\documentclass{elsart}
\usepackage{natbib}

\begin{document}
\begin{frontmatter}

\title{A Role of the Boundary Shear Layer \\
       in Modeling of Large Scale Jets\thanksref{grant}}

\author{\L . Stawarz} and
\author{M. Ostrowski}
\address{Obserwatorium Astronomiczne, Uniwersytet Jagillonski,
         ul. Orla 171, 30-244, Krak\'{o}w, Poland}
\thanks[grant]{Work supported by KBN through the grant BP 258/P03/99/17}

\begin{abstract}
We discuss a role of the boundary shear layer of large scale jets for their 
observed multiwavelength emission. We consider a simple mechanism of a turbulent 
particle acceleration acting within such regions as an alternative mechanism 
with respect to the standard approaches involving internal shocks in the jet. 
We show, that under simple assumptions the boundary layer acceleration can 
provide very high energy electrons possibly generating the X-ray emission 
observed by {\it Chandra} from some large scale jets in powerful radiogalaxies.
\end{abstract}

\end{frontmatter}

\section{Introduction: the aim of the study}

Large scale extragalactic jets, extending from a few to a few hundreds of
kiloparsecs from active galactic nuclei (AGNs), were frequently studied at
radio frequencies. Recently, {\it Hubble} and {\it Chandra} telescopes gathered 
new detailed information about optical and X-ray emission of some of these
objects. For a few hundreds of radio jets we know, only $\sim$ 20 are observed at
optical frequencies. Most of them are short and faint, with only a few exceptions
allowing for detailed spectroscopic and morphological studies, like e.g. 3C 273 
\citep{jes01}. Surprisingly, the large scale jets can be very prominent at X-rays
in many different types of radio-loud AGNs. Up to now, more than 30 jets were
detected by {\it Chandra} at $1 - 10$ keV energy range, although a nature of
this emission is still under debate. 

The {\it Hubble} observations indicate with no doubt a synchrotron nature of 
the optical emission. The first conclusion from the optical maps is a requirement
of electron reacceleration within the whole jet volume, as their radiative 
lifetimes are usually much shorter than the time required for light to travel
between the successive shocks. Sub-equipartition magnetic field and highly 
relativistic bulk velocities alone cannot remove this problem \citep{jes01}. 
Also, a spectral character of the radio-to-optical continuum, if carefully analyzed, 
is not consistent with simple versions of the shock-in-a-knot models. The X-ray 
observations are even more puzzling, because in many cases it is not clear if the 
detected keV photons -- probably non-thermal in nature -- result from synchrotron 
or inverse-Compton emission. In any case, in order to explain the observed high
luminosities and spectra of the {\it Chandra} jets, one has to invoke more or less 
extreme conditions, like large beaming factors or presence of very high energy 
electrons \citep{har01}. Once again, details of the electron acceleration processes 
responsible for the observed X-ray emission are not clear.\\
On the other hand, radio and optical observations reveal also a complex jet 
spatial structure indicating a spine - boundary shear layer morphology. As 
emphasized below, such boundary shear layers are prevailed sites of particle 
acceleration, especially in cases of relativistic flows, and therefore should be 
seriously considered as an option/addition to the shock-in-a-knot models in 
studying the jet multiwavelength emission. In fact, contribution of the shear layer 
to the jet radiative output results not only from the involved electron acceleration, 
but also from kinematic effects influencing the composed relativistic jet spectrum.
Recent optical and X-ray observations give us an important insight into the mentioned
processes and effects.

\section{Sheared jets: electron acceleration}

As pointed by \citet{dey86}, interaction between the jet matter and the surrounding
medium result in formation of the shear layer at the jet boundary. Such boundary 
regions are likely to be highly turbulent because of their very high Reynold numbers. 
Therefore, stochastic particle acceleration acting thereby seems to be also inevitable.
Numerical simulations confirm presence of the turbulent layer with a velocity shear
surrounding fast central spine of the jet \citep{alo99}, although the exact nature of
the shear regions is still hardly known. With no doubt, they play an important role
in stabilizing the flow with respect to the Kelvin-Helmholtz (KH) instabilities
\citep{bir91}. They are the places where the large-scale KH instabilities form and
cascade to shorter wavelengths, which dissipate turbulence energy by resonant
interactions with thermal and relativistic particles \citep{eil82}. Such stochastic
interactions can be additionaly influenced by weak oblique shocks developed within
the boundary shear regions, as suggested once again by numerical simulations.
In addition, a presence of the velocity gradient can also modify the particle
energy distribution due to effects of a `cosmic ray viscosity', discussed e.g. by
\citet{ost00}. Thus, modeling of the electron acceleration processes acting within
the boundary shear layers of large scale jets is difficult and requires several
assumptions about the boundary layer internal structure. One of these assumptions
refers to the magnetic field configuration and intensity. Both observations and
theoretical considerations suggest that the magnetic field in the boundary regions
is parallel on average to the jet axis due to shearing effects. One should expect
that a strong velocity shear may influence not only the magnetic field configuration,
but also its intensity by a dynamo process \citep{urp02}.

An order-of-magnitude analysis of time scales connected with electron momentum and
spatial diffusion in a boundary layer medium shows, that for typical parameters of
the large scale relativistic jets in powerful radio sources the boundary layer
electrons undergo mainly second-order Fermi acceleration, and that the time scale
for electron escape from the acceleration region is extremely long as compared to
the radiative losses time scale \citep{sta02a}. As a result, electrons form a flat
power-law energy distribution with a harder high energy component modelled by us
as a pile-up bump at the maximum energy \citep{ost00}. Under the condition of
continuous and {\it efficient} electron acceleration acting within the whole highly
turbulent boundary layer volume, involving large amplitude MHD turbulence leading
to the scattering mean free path comparable to the electron gyroradius, a balance
between acceleration and radiative losses allow for electron maximum energy,
typically, $E_{eq} \sim 10^8 \, m_e c^2$. One should note, however, that formation
of the terminal hard component in the electron energy spectrum can be a non-stationary
process \citep{sta02b}. In principle, its normalization grows with time for the
considered continuous particle injection, while its spectral width increases over
larger and larger energy range due to momentum diffusion. In \citet{sta02a} we
modeled the pile-up bump as a monoenergetic peak with total energy density
limited by the equipartition with the magnetic field. This requires further
studies of the electron energy evolution \citep{sta03}. However, we would like to
stress out, that the pile-up effects forming a final stationary hard spectral
component at the highest electron energies seem to be inevitable in the considered
case of the boundary layer acceleration.

\section{Conclusions: boundary shear layer emission}

It is known, that emission from the shear boundary layer can decrease
jet-counterjet radio brightness asymmetry, influencing estimations of the jet
bulk Lorentz factors \citep{kom90}. Also, because of the kinematic effects,
different radiation fields can dominate the inverse-Compton emission of the 
boundary layer electrons as compared to the spine electrons, affecting the 
observed jet high energy radiation \citep{cel01}. However, an important effect 
of the boundary shear layer on the jet radiative output is due to stochastic 
and continuous electron acceleration acting within the whole boundary layer 
volume, most likely resulting in effective accumulation of the radiating 
electrons around the maximum energy $E_{eq}$. For typical large scale jet 
parameters, synchrotron radiation of the electrons with $E \ll E_{eq}$ can 
account for almost constant along the jet radio-to-optical continuum, while 
the electrons with $E \sim E_{eq}$ can be responsible -- at least for some 
sources -- for the relatively strong X-ray emission detected by {\it Chandra}
\citep{sta02a}. Combination of the effects connected with spectral pile-ups and
relativistic beaming in a medium with velocity shear, as well as an interplay
between the stochastic and the shock acceleration, can possibly explain spectral 
variety and complexity of the {\it Chandra} jets.\\
In our simple model we in fact consider two distinct relativistic electron
populations, which differ because of the spatial location (knots -- jet edges),
nature of the acceleration (regular -- stochastic) and kinematic effects
involved (fast central spine -- shear boundary layer). However, additional
processes can also lead to several complications of such simple two-population
model, like for instance formation of oblique shocks within the boundary region
\citep{bic82} or a turbulent mixing of the spine and the boundary layer medium.
Whatever the case is, the efficient particle acceleration taking place at the
jet boundary can influence the observed jet radiative properties. By its nature,
it is also connected with the issue of jet internal structure and jet stability
in relates to the MHD instabilities excited at the jet surface. Thus, to
understand multiwavelength emission of the jets it is important to study a role
of its boundary shear layer, and the acceleration processes acting thereby.

\end{document}